\documentclass[conference]{IEEEtran}

\IEEEoverridecommandlockouts                              %

\usepackage{graphics} %
\usepackage{epsfig} %
\usepackage{times} %
\usepackage[cmex10]{amsmath} %
\usepackage{bm}
\usepackage{algorithmic}
\usepackage{algorithm}
\usepackage{caption} %
\usepackage{subcaption} %
\usepackage[dvipsnames]{xcolor}
\usepackage{multirow}
\usepackage{cite}
\usepackage[hyphens]{url}

\renewcommand{\baselinestretch}{0.98}

\title{
Co-optimization of power line shutoff and restoration
under high wildfire ignition risk\\
\thanks{This work was supported by the U.S. National Science Foundation ASCENT program under award 2132904.}
}

\author{\IEEEauthorblockN{Noah Rhodes* and Line A. Roald}
\IEEEauthorblockA{\textit{Department of Electrical and Computer Engineering}, 
\textit{University of Wisconsin-Madison}\\
Madison, WI United States \\
*nrhodes@wisc.edu}
}

\newfloat{model}{thp}{lop}
\floatname{model}{Model}

\begin{document}

\maketitle
\thispagestyle{empty}
\pagestyle{empty}

\begin{abstract}
Electric power infrastructure has ignited several of the most destructive wildfires in recent history. Preemptive power shutoffs are an effective tool to mitigate the risk of ignitions from power lines, but at the same time can cause widespread power outages. 
This work proposes a mathematical optimization problem to help utilities %
decide where and when to implement these shutoffs, as well as how to most efficiently restore power once the wildfire risk is lower. Specifically, our model co-optimizes the power shutoff (considering both wildfire risk reduction and power outages) as well as the post-event restoration efforts given constraints related to inspection and energization of lines, and is implemented as a rolling horizon optimization problem that is resolved whenever new forecasts of load and wildfire risk become available. We demonstrate our method on the IEEE RTS-GMLC test case using real wildfire risk data and forecasts from US Geological Survey, and investigate the sensitivity of the results to the forecast quality, decision horizon and system restoration budget.  The software implementation is available in the open source software package \emph{PowerModelsWildfire.jl}.
\end{abstract}

\begin{IEEEkeywords}
Wildfire risk mitigation, preemptive power shutoffs, grid restoration, 
mixed-integer optimization.
\end{IEEEkeywords}

\section{INTRODUCTION} \label{sec:introduction}
Wildfire ignitions caused by electrical equipment are an increasing concern for power grid operators.  
In Australia,
electrical infrastructure accounts for 30\% of bushfire ignitions during droughts and heat-waves~\cite{2009_victorian}. 
The state firefighting organization in California
reports that between 2015-2020, wildfires ignited by electrical equipment were responsible for more than 70\% of damages (\$17.5 billion) despite accounting for only 10\% of ignitions
\cite{noauthor_stats_nodate}.
Given these insights, focus on wildfire-grid interactions has expanded from almost exclusively considering how the grid is impacted by fires (discussed in e.g. \cite{7995099, CHOOBINEH201520, mohagheghi2015optimal, moutis2022pmu, nazemi2022powering}) to also consider how to reduce the risk of wildfire ignitions. Refs. \cite{8768218, 8767948, arab2021three, muhs2020wildfire} review approaches to reduce the risk of ignitions, %
many of which are already being implemented or planned by utilities. For example, Pacific Gas \& Electric (PG\&E) in California 
plans to underground 10,000 miles of distribution lines over the course of more than a decade at a cost of \$20 billion \cite{noauthor_pge_nodate}, along with other measures such as 
more frequent vegetation clearing. %
Unfortunately, the significant cost and need for qualified crews and equipment means that these risk reduction approaches take time to implement.

Meanwhile, power system operators rely on short-term measures to reduce risk, including
changes to the protection system settings %
\cite{7995099} or Public Safety Power Shutoffs (PSPS), which de-energize grid equipment during high wildfire risk events. %
While highly effective in reducing wildfire risk \cite{noauthor_psps_nodate}, 
PSPS also cause
large scale power outages that may last for days \cite{noauthor_utility_nodate} with significant economic and health impacts. 
To address the question of how to balance the benefits of wildfire risk reduction with the impacts of power outages, \cite{rhodes2020balancing} proposed a framework to optimally balance wildfire risk reduction and power outage sizes when deciding which lines to shut off. 
Other approaches focus on accelerating the solution time to enable efficient planning of optimal PSPS on large networks in real time, including data-driven methods to plan PSPS by training machine learning models with optimization problems results \cite{hong2022data, umunnakwe2022data, bayani2022quantifying}, or using a dynamic programming approach to optimize a PSPS \cite{lesage2022optimally}.  %
Studies on how to reduce the impact of and need for PSPS include \cite{kody2022optimizing}, which proposes an optimal investment model for installation of batteries and under-grounding power lines, and \cite{yang2022resilient}, which studies how microgrids can improve resilience against unplanned outages due to events like wildfires, and \cite{astudillo2022managing}, which expands on time-varying risk and operational factors such as energy storage for temporal load-shifting. Other related research considers extensions to distribution grids \cite{gorka2022efficient}, %
consideration of fairness of repeated PSPS events \cite{kody2022sharing}, and improved forecasting of wildfire ignition risk \cite{yao2022predicting, bayani2022quantifying}.

Although the above methods may help %
identify locations where de-energization is needed, they do not consider the post-event restoration process. %
Before a line can be re-energized, it must be inspected by utility crews to ensure that no damage that could cause ignitions has occurred.
Thus, while de-energization of a line can happen instantaneously, the restoration process can take hours or days \cite{pgeOct2020report}. 
To balance power outages and ignition risk, we therefore need to consider the size and duration of power outages both during the initial shut-down and throughout the restoration process.

In this paper, we aim to address this gap.  Our first and main contribution is to develop the \emph{multi-period optimal power shutoff and re-energization} (MOPSAR) problem, a mixed-integer optimization model which co-optimizes power shutoffs and re-energization. %
The formulation leverages prior work on balancing wildfire risk and power outages for a single time period \cite{rhodes2020balancing} and post-disaster restoration planning \cite{rhodes2021powermodelsrestoration}, but significantly extends these (and other) existing formulations in several ways. First, our model is the first to integrate both power shutoffs and re-energization in one model. Second, %
we solve the MOPSAR problem as a \emph{rolling horizon} problem which is rerun daily with updated wildfire risk forecasts, which allows us to consider evolving wildfire risk over a prolonged time horizon and mitigate the impact of forecast errors. Third, we include a new penalty on shutoffs of low-risk lines, as such power shutoffs may increase the operational vulnerability of the network during a shutoff.

Our second contribution is to use our improved problem formulation to investigate important aspects of the problem. We assess how line inspection constraints impact both wildfire risk and power outage sizes, how using a longer forecast horizon impacts the solution quality and computational time, and how the choice of power shutoff penalty impacts the trade-off between grid vulnerability and wildfire risk.

The remainder of the paper is organized as follows. 
Section \ref{sec:formulation} introduces the problem formulation.  Section \ref{sec:experiment} presents the case study and problem setup, and analysis of  the results.  Section \ref{sec:conclusion} concludes the work. 

\section{Modeling and Problem Formulation} \label{sec:formulation}

In this section, we formulate the Multi-period Optimal Power Shutoff And Restoration (MOPSAR) problem %
that provides an optimized schedule for both power line shutoffs and restoration. 
It is solved daily in a rolling horizon framework as new risk forecast data is made available each day.

\subsection{Rolling Horizon Formulation}
We consider a power grid with a set of buses $\mathcal{B}$, lines $\mathcal{L}$, generators $\mathcal{G}$ and loads $\mathcal{D}$. This grid has elevated wildfire risk, and the operator is planning a public safety power shutoff to mitigate the wildfire threat.  The operator must make decisions on the power lines they may shutdown, customers they may disconnect, and how to re-energize the grid.  Their goal is to reduce wildfire risk while minimizing the size and duration of customer power outages and maintaining grid reliability.

Wildfire risk depends on ambient conditions such as vegetation cover, humidity and wind speed. %
The Wildland Fire Potential Index (WFPI) \cite{wfpi} provides wildfire risk data for the contiguous United States, is released daily and includes current data along with 7-day forecasts. %
Because it takes time to restore a power line after it has been shut off, we need to take the forecasted risk into account when deciding on an optimal schedule for power shutoffs and restoration. Furthermore, because the wildfire risk forecasts evolve and become more accurate with time, it makes sense to update the schedule daily as new information becomes available. 
We therefore formulate our problem as a rolling horizon problem. 

For a given day $T$, a Multi-period Optimal Power Shutoff And Restoration (MOPSAR) problem, shown in Model \ref{model:mops}, is solved. This problem considers the status of the power lines at the beginning of day $T$ (i.e. whether a line is energized or not) as well as grid constraints, and current and forecasted wildfire risk for all timesteps $t\in\mathcal{T}$ where
$\mathcal{T}=\{T,..., T+\boldsymbol{H}\}$ includes the current day and all days in the forecast horizon $\boldsymbol{H}$. 
Once a solution is obtained, the power shutoff and restoration actions for the first day $t=T$ are then implemented on the power grid. The next day, we repeat the optimization process with $T \leftarrow T+1$, incorporating the updated power line statuses based on the solution from the previous day and updated wildfire risk information.  
We next present the mathematical formulation of the MOPSAR problem. %

\subsection{Objective function}
The system operator is pursuing three different objectives, namely maximizing load, minimizing wildfire risk and maintaining grid reliability. We describe each of these below. 

\subsubsection{Load served}
The load $D_{Served}$ is calculated as
\begin{equation}
    {D_{Served}} = \sum_{ t \in \mathcal{T}} \sum_{d\in\mathcal{D}} x_{d t} w_{d t} \boldsymbol{P}^D_{d t}. \label{eq:dtot}
\end{equation}

Here, $\boldsymbol{P}^D_{d t}$ is a parameter expressing the total demand for electricity from load $d$ at time period $t$ and $w_{d t}$ is a weight used to express increased priority for certain loads (e.g. a hospital may have a higher weight). We assume that load can be continuously shed, with the continuous decision variable $x_{d t}$ representing the percentage of load that is served. 
The total demand served $D_{Served}$ is obtained by summing over all loads $d \in \mathcal{D}$ and all time periods $t\in\mathcal{T}$. 

\subsubsection{Wildfire risk}
We express the total risk of wildfire ignitions based on the wildfire risk associated with each transmission line in the network, i.e. 
\begin{equation}
    {R_{Fire}} = \sum_{ t \in \mathcal{T}} \sum_{ij\in\mathcal{L}}z_{ij t} \boldsymbol{R}_{ij t} \label{eq:risktot}
\end{equation}
Here, $\boldsymbol{R}_{ijt}$ represents the risk of a wildfire ignition from line $ij$ at time $t$ if the line is energized (we discuss how to obtain these risk values in the case study). The binary decision variable $z_{ijt}$ represents the status of the line, with $z_{ijt}=1$ indicating that it is on and $z_{ijt}=0$ indicating that it is off. 
This formulation assumes that by choosing to de-energize the line, i.e. 
$z_{ijt}=0$, we reduce the wildfire risk of line $ij$ at time $t$ to zero.

\subsubsection{Grid Vulnerability}
Because grid operational security benefits from redundant transmission paths,
we would like to keep low risk lines in operation even if they do not contribute to serving more load. %
To incorporate the effect of power shutoff on grid operational security, we introduce a penalty on power shutoffs,
\begin{equation}
    {V}_{system} = \sum_{ t \in \mathcal{T}} \sum_{ij\in\mathcal{L}} (1-z_{ij t})\boldsymbol{V} \label{eq:off_threshold}
\end{equation}
Here, $\boldsymbol{V}$ represents the increased vulnerability of the grid associated with turning off an individual line and $V_{System}$ represents the total vulnerability of the network. 

Next, we combine  $R_{Fire}$ and $V_{System}$ in a single term,
\begin{align}
    R_{Fire}-V_{System} = |\mathcal{L}||\mathcal{T}|\boldsymbol{V} + \sum_{ t \in \mathcal{T}} \sum_{ij\in\mathcal{L}}z_{ij t} (\boldsymbol{R}_{ijt}-\boldsymbol{V})
\end{align}
This expression highlights that $\boldsymbol{V}$ represents the minimum wildfire risk for which it is beneficial to disable a line, and we will hence refer to $\boldsymbol{V}$ as the \emph{risk threshold}\footnote{We recognize that expressing grid vulnerability on a line-by-line basis is less comprehensive and meaningful than using other metrics such as N-1 security. However, it does allow us to express a preference for keeping low risk lines in operation. A more in-depth modeling and investigation of grid vulnerability and operational security in the context of high wildfire risk is left for future work.}. When $R_{ijt} > \boldsymbol{V}$, it is beneficial to turn of line $(i,j)$ to reduce wildfire risk in the system. When $R_{ijt} < \boldsymbol{V}$, the wildfire risk of line $(i,j)$ is not high enough to have a benefit of disabling the line. %

Given the above modeling considerations, we formulate the objective function \eqref{eq:objective}. The objective function uses the total load $\boldsymbol{D}_{Tot}$ and the total wildfire risk before a power shutoff $\boldsymbol{R}_{Tot}$ as normalization factors, with $\boldsymbol{D}_{Tot}, \boldsymbol{R}_{Tot}$ defined as 
\begin{align}
    \boldsymbol{D}_{Tot}=\sum_{ t \in \mathcal{T}} \sum_{d\in\mathcal{D}} \boldsymbol{P}^D_{d t}, \quad \boldsymbol{R}_{Tot}= \sum_{ t \in \mathcal{T}} \sum_{ij\in\mathcal{L}} \boldsymbol{R}_{ij t}
\end{align}
With this normalization, the objective expresses the percentage of load and wildfire risk after implementation of the PSPS, with $0\leq D_{Served}/\boldsymbol{D}_{Tot}\leq 1$ and $0 \leq R_{Fire}/\boldsymbol{R}_{Tot} \leq 1$.

The trade-off parameter $\alpha \in [0,1]$ allows us to express a preference for serving load or mitigating wildfire risk. The preference for mitigating wildfire risk vs limiting grid vulnerability is given by our choice of the risk threshold $V$. 

\subsection{Restoration Constraints}
An important and novel aspect of our formulation is the consideration of a limited \emph{restoration budget}, i.e. a limited capacity to inspect and restore power lines. 
The limits on how many miles of lines can be restored in each time period is described by constraints \eqref{eq:resto_indicator}-\eqref{eq:resto_limit}.
Constraints \eqref{eq:resto_indicator} set the indicator variable for restoration $y^L_{i j t}$ to 1 if a line is off in the previous period $z^L_{i j t-1}=0$ and on in the current period $z^L_{i j t}=1$. These logical constraints are implemented in the optimization problem by \eqref{eq:logic_constraints}, where the three inequalites form the logical \emph{negation} of $z^L_{ij t-1}$ as well as the \emph{and} operation, 
\begin{subequations}
\vspace{-3pt}
    \begin{align}
        & y^L_{ijt} \le (1-z^L_{ij t-1}) \\
        & y^L_{ijt} \le z^L_{ijt} \\
        & y^L_{ijt} \ge (1-z^L_{ijt-1}) + z^L_{ijt}-1
    \end{align}
    \label{eq:logic_constraints}
\end{subequations}
Constraint \eqref{eq:resto_indicator_0} implements the same constraint for the initial condition parameter $\boldsymbol{z}^L_{i j 0}$ of the line in the first period. 
Constraint \eqref{eq:resto_limit} limits the amount of restoration that can occur in a single period according to the restoration budget $\boldsymbol{Y}_t$.  The effort required to restore a line scales with length $\boldsymbol{\ell}_{i j}$, %
and the parameter $\boldsymbol{Y}_t$ thus represents the total length of lines that the utility is able to inspect in time step $t$.
Finally, \eqref{eq:bin_var} enforces that  $z^L_{ij t}$ and $y^L_{i j t}$ take on binary values.

\begin{model}[t]
    \centering
    \begin{subequations}
        \vspace{-0.2cm}
        \begin{flalign*}
            &  \quad \mbox{\bf variables: ($\forall t \in \mathcal{T}$)} &\\
            & \quad P_{g t}^G  \forall g \! \in\! \mathcal{G},  ~
                        P_{ij t}^{L}  \forall (i,j) \!\in\! \mathcal{L},   ~
                        \theta_{i t} \forall i \!\in\! \mathcal{B}, \\
            & \quad
                        x^D_{d t}  \forall d \!\in\! \mathcal{D}, ~
                        z^L_{ijt}  \forall (i,j) \!\in\! \mathcal{L}, ~
                        y^L_{ijt}  \forall (i,j) \!\in\! \mathcal{L} &
        \end{flalign*}
        \vspace{-1.5em}
        \begin{align}
        & \mbox{\bf maximize:} && \!\!\!\!\! (1-\boldsymbol{\alpha})\frac{D_{Served}}{\boldsymbol{D}_{Tot}} \!-\! \boldsymbol{\alpha}\left( \frac{{R_{Fire} \!-\! V_{System}}}{\boldsymbol{R}_{Tot}}\right)
        \label{eq:objective}
        \end{align}
        \vspace{-1.5em}
        \begin{align}
        &\mbox{\bf subject to } \forall t \in \mathcal{T}: \nonumber \\
        & {\textstyle{D_{Served}} = \sum_{ t \in \mathcal{T}} \sum_{d\in\mathcal{D}} x_{d t} w_{d t} \boldsymbol{P}^D_{d t}} \\
        & {\textstyle{R_{Fire}} = \sum_{ t \in \mathcal{T}} \sum_{ij\in\mathcal{L}}z_{ij t} \boldsymbol{R}_{ij t}} \\
        & {\textstyle{V}_{system} = \sum_{ t \in \mathcal{T}} \sum_{ij\in\mathcal{L}} (1-z_{ij t})\boldsymbol{V}}  \\[+2pt]
        & y^L_{i j t} = (\lnot z^L_{i j t-1}) \wedge z^L_{i j t} && \mkern-90mu \forall (i,j) \in \mathcal{L}, ~ \text{for } t \neq \boldsymbol{T} \label{eq:resto_indicator}\\
        & y^L_{i j 1} = (\lnot \boldsymbol{z}^L_{i j 0}) \wedge z^L_{i j t} && \mkern-90mu \forall (i,j) \in \mathcal{L}, ~ \text{for } t = \boldsymbol{T}  \label{eq:resto_indicator_0}\\
        & \sum_{(i j)\in L} y^L_{i j t} \boldsymbol{\ell}_{i j} \leq \boldsymbol{Y}_t \label{eq:resto_limit} \\
        & z^L_{ij t} \in \{0,1\}, \quad y^L_{i j t} \in \{0,1\} && \forall (i,j) \in \mathcal{L} \label{eq:bin_var} \\
        & \sum_{g\in\mathcal{G}_i} P_{g t}^G +\!\!\! \sum_{(i,j)\in\mathcal{L}_i} \!\!\!P_{ t}^L -\!\!\! \sum_{d\in\mathcal{D}_i} x^D_{d t} \boldsymbol{P}^D_{d t} = 0 && \forall i \in \mathcal{B}
        \label{eq:power_balance} \\
        & 0\le x^D_{d t} \le 1 && \forall d \in \mathcal{D} \label{eq:load_bounds} \\
        & 0 \le P_{g t}^G \le  \overline{\boldsymbol{P^G}_{g t}} && \forall g \in \mathcal{G}
        \label{eq:gen_limits}\\
        &-\overline{\boldsymbol{P^L}_{ij}}z^L_{ijt} \le  P_{ijt}^{L} \le \overline{\boldsymbol{P^L}_{ij}}z^L_{ijt} && \forall (i,j) \in {\mathcal{L}}
        \label{eq:thermal_limit_d}\\
        & P_{ij t}^{L} \le -\boldsymbol{b}_{ij} (\theta_{i t} - \theta_{j t}) + \overline{\boldsymbol{\theta^{\Delta}}_{ij}} (1-z^L_{ij t}) && \forall (i,j) \in \mathcal{L}
        \label{eq:flow_limit1_d} \\
        & P_{ij t}^{L} \ge -\boldsymbol{b}_{ij} (\theta_{i t} - \theta_{j t}) + \underline{\boldsymbol{\theta^{\Delta}}_{ij}} (1-z^L_{ij t}) &&  \forall (i,j) \in \mathcal{L} 
        \label{eq:flow_limit2_d}
        \end{align}
    \end{subequations}
    \caption{Multi-period Optimal Power Shutoff And Restoration (MOPSAR)}
    \label{model:mops}
    \vspace{-0.2cm}
\end{model}

\subsection{Power Flow Constraints}
To model how de-energization and restoration of lines impact the amount of electricity served to customers, we need to model the power flows in the system.
Equations \eqref{eq:power_balance}-\eqref{eq:flow_limit2_d} %
represents the DC power flow in each time period, while accounting for load shedding and energization status of lines. 

Nodal power balance is enforced by \eqref{eq:power_balance} where total generation $P^G_{g t}$, transmission line power $P^L_{ijt}$, and load served $x^D_{d t} \boldsymbol{P}^D_{d t}$ must sum to zero at each bus. The sets $\mathcal{G}_i$, $\mathcal{L}_i$, $\mathcal{D}_i$ represent the sets of generators, lines and load demands at bus $i$. Eq. \eqref{eq:load_bounds} ensures that the fraction of load served for each load $x^D_{dt}$ is constrained to be within 0 and 1.
Eq. \eqref{eq:gen_limits} ensures that the power $P^G_{g t}$ from each generator $g$ is non-negative and below the upper limit $\overline{\boldsymbol{P^G}_{g t}}$, which  may vary with the time period $t$ for renewable energy sources, based on the forecasted maximum output. While the problem formulation can support non-zero lower bounds on generation by including additional binary variables for generator on/off status, we remove the unit commitment aspect in this paper for simplicity. 

Constraint \eqref{eq:thermal_limit_d} keeps the power flow $P^L_{ijt}$ on the line from $i$ to $j$ between the power limits $-\overline{\boldsymbol{P^L}_{ij}}z^L_{ijt}$ and $\overline{\boldsymbol{P^L}_{ij}}z^L_{ijt}$ when the line is energized, i.e. $z^L_{ijt}=1$.  When a line is de-energized, $z^L_{ijt}=0$, the power flow across the line is 0.  
Equations \eqref{eq:flow_limit1_d}, \eqref{eq:flow_limit2_d} define the line power flow, while accounting for line energization status. When a line is energized, $z^L_{ijt}=1$, these constraints reduce to the ordinary linearized DC power flow,
\begin{equation}
   P_{ij t}^{L} = -\boldsymbol{b}_{ij} (\theta_{i t} - \theta_{j t}), \\
\end{equation}
where $\boldsymbol{b}_{ij}$ is the line suseptance, and $\theta_{it}$ and $\theta_{jt}$ is the bus voltage angle for bus $i$ at time $t$. 
When the line is de-energized, $z_{ijt}=0$, the power flow is decoupled from the voltage angle difference through the big-M values $\overline{\boldsymbol{\theta^{\Delta}}_{ij}}$ and $\underline{\boldsymbol{\theta^{\Delta}}_{ij}}$, which can be calculated as in \cite{Burak2016}. This allows the power flow to be set to $0$ in \eqref{eq:thermal_limit_d} without constraining the voltage angle differences.

\section{Case Study} \label{sec:experiment}
We next demonstrate the efficacy of our proposed model and study the impact of %
the forecast horizon, restoration budget, and the risk threshold on solutions. %

\subsection{Case study setup}
We first describe our implementation and the data used. %

\subsubsection{Implementation}
The optimization model is available in the open source package \emph{PowerModelsWildfire.jl} \cite{rhodes2020balancing}, implemented in the Julia language \cite{julia}. We use the Gurobi v9.1 optimization solver \cite{gurobi} with an optimality gap of 0.01\%.

\subsubsection{Test System}
We base our case study on the RTS-GMLC \cite{RTSGMLC} system.  This synthetic test system has geographic coordinates located in southern California, a region which has been affected by Public Safety Power Shutoffs. The system has one year of hourly load and renewable energy profiles, and we use the data from the wildfire season in October and November. Because we use wildfire risk data with one daily value, we %
select the hour with highest level of system load, where the demand is the most difficult to serve, from each day to represent the power flows. 
The RTS-GMLC system has relatively high power limits on the transmission lines.
To obtain a more interesting case with more realistic congestion, we increased the active power demand using the API method in \cite{pglib}, which finds the maximum amount all loads can be scaled to assuming generation is unconstrained.  
This method  scaled each load in the network by a factor of 2.14, and we scaled the  capacity of each generator by the same amount. %
We use a uniform load priority weight $w_{dt}=1$.

\subsubsection{Wildfire risk data}
Wildfire risk data was obtained from the United States Geological Survey's  (USGS) Wildland Fire Potential Index (WFPI) \cite{wfpi}. %
The WFPI incorporates fuel models and forecasts for precipitation, dry bulb temperature and wind speeds to create a daily forecast of the WFPI for up to 7 days.  
For each line and each day, we calculate the wildfire risk as the highest WFPI value along the line \cite{taylor2022framework}. 

\begin{figure}[t]
    \centering
    \begin{subfigure}{0.9\columnwidth}
        \centering
        \includegraphics[width=1.0\columnwidth]{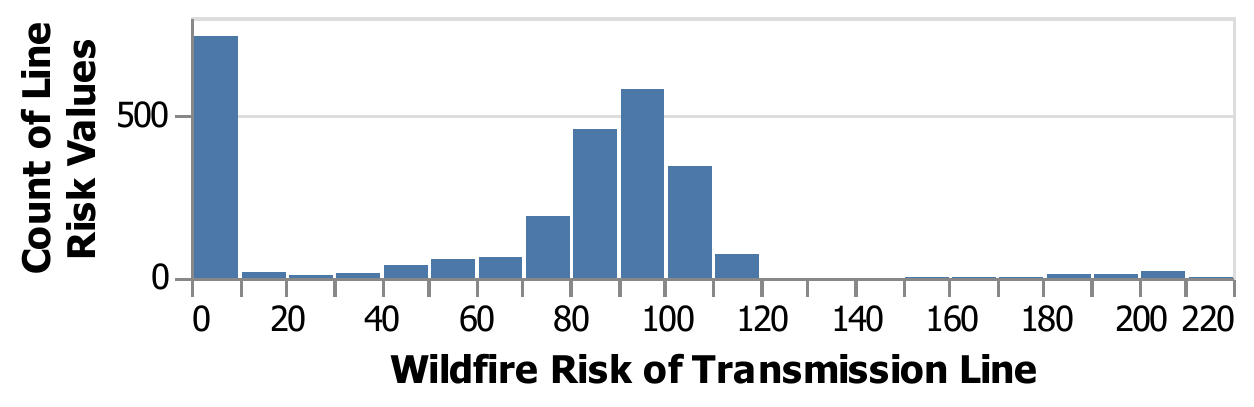}
        \caption{\small \textbf{Power Line Risk}
        }
        \label{fig:line_risk_bar}
    \end{subfigure}
    \begin{subfigure}{0.9\columnwidth}
        \centering
        \includegraphics[width=1.0\columnwidth]{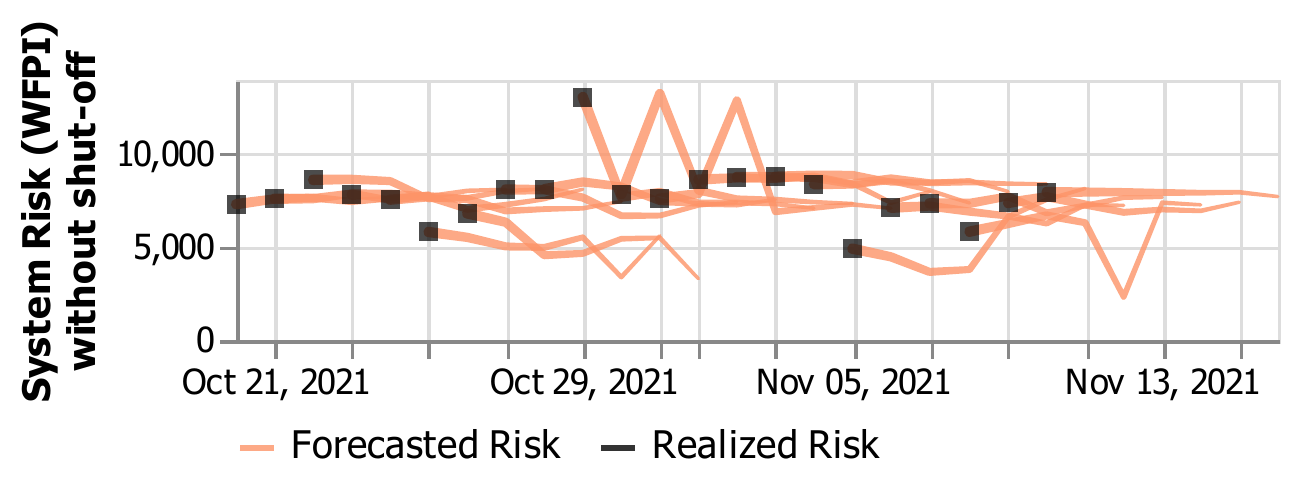}
        \caption{\small 
        \textbf{Risk Forecasts}
        }
        \label{fig:forecasted_risk}
    \end{subfigure}
    \caption{\small System Wildfire Risk Data: Top: Histogram of the wildfire risk of individual transmission lines, based on all realized wildfire risk observed from Oct 20th to Nov 10th. Bottom: Realized (black) and forecasted (orange lines) wildfire risk for the overall system, calculated as the sum of all wildfire risk values across all lines.} %
    \label{fig:risk_data}
    \vspace{-10pt}
\end{figure}

Figure \ref{fig:line_risk_bar} shows a histogram of all the realized wildfire risk values (i.e. not including forecasted data) for all lines from Oct 20th to Nov 10th 2021. During the study period many lines have either zero risk or risk in the range between 70-120, while a small number of lines have extreme risk $> 160$.

Figure \ref{fig:forecasted_risk} shows the total system risk before any power shutoffs (i.e., the sum of all wildfire risk values for all lines, assuming all lines are energized) based on the forecasted (in orange) and realized (in black) wildfire risk values. %
There are substantial forecast errors, demonstrating the importance of resolving the problem as new data becomes available. %

\subsubsection{Baseline Parameters}
We set $\alpha=0.7$ as we found this to provide a trade-off with significant risk reduction and minimal load shed under this specific loading and wildfire risk scenario, and allows us to analyse the impact of new parameters in MOPSAR problem.  
For analysis on the impact of changing $\alpha$, refer to \cite{rhodes2020balancing}. 
Unless otherwise specified, we set the restoration budget to $\boldsymbol{Y}_t=75$ miles/day, the forecast horizon to $4$ days and the risk threshold to $\boldsymbol{V}=100$ WFPI.

\begin{figure}[t]
    \centering
    \begin{subfigure}{0.85\columnwidth}
        \centering
        \includegraphics[width=1.0\columnwidth]{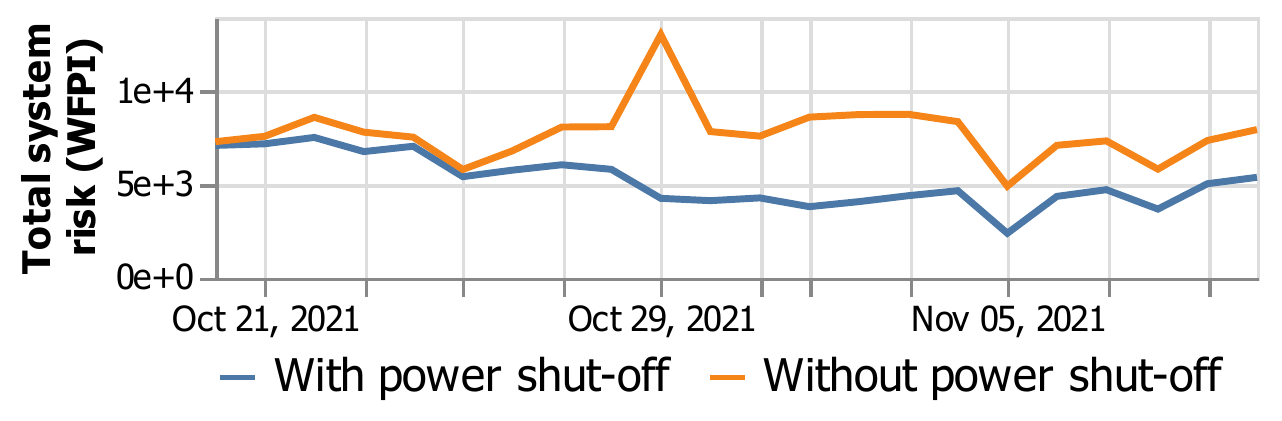} 
     \end{subfigure}
     \begin{subfigure}{0.9\columnwidth}
        \centering
        \includegraphics[width=1.0\columnwidth]{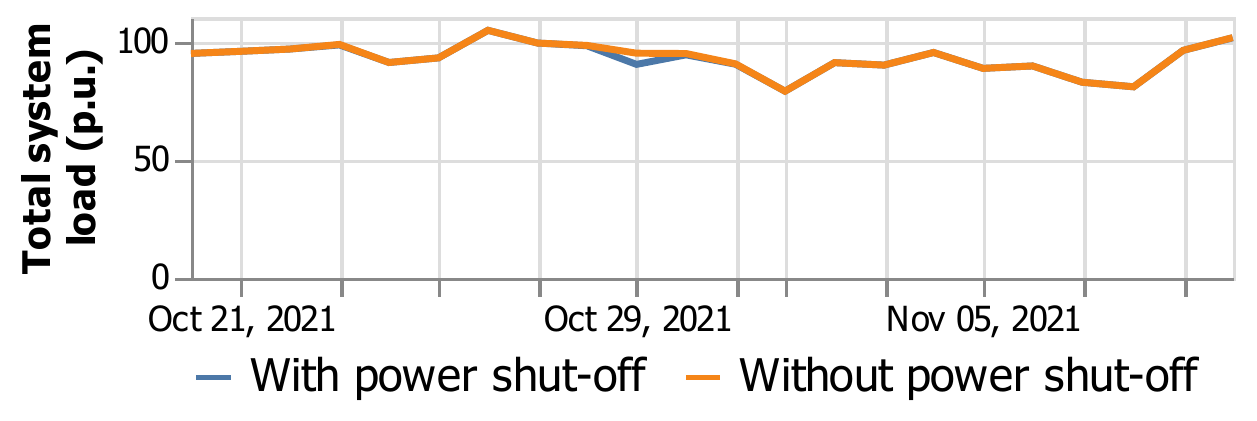} 
    \end{subfigure}
    \caption{\small Baseline MOPSAR solutions.  Top: Comparison of wildfire risk with and without power shutoffs. Bottom: Total system load served with and without power shutoffs.}
     \label{fig:system_plot}
     \vspace{-10pt}
\end{figure} 

\begin{figure*}
    \centering
    \includegraphics[width=0.85\textwidth]{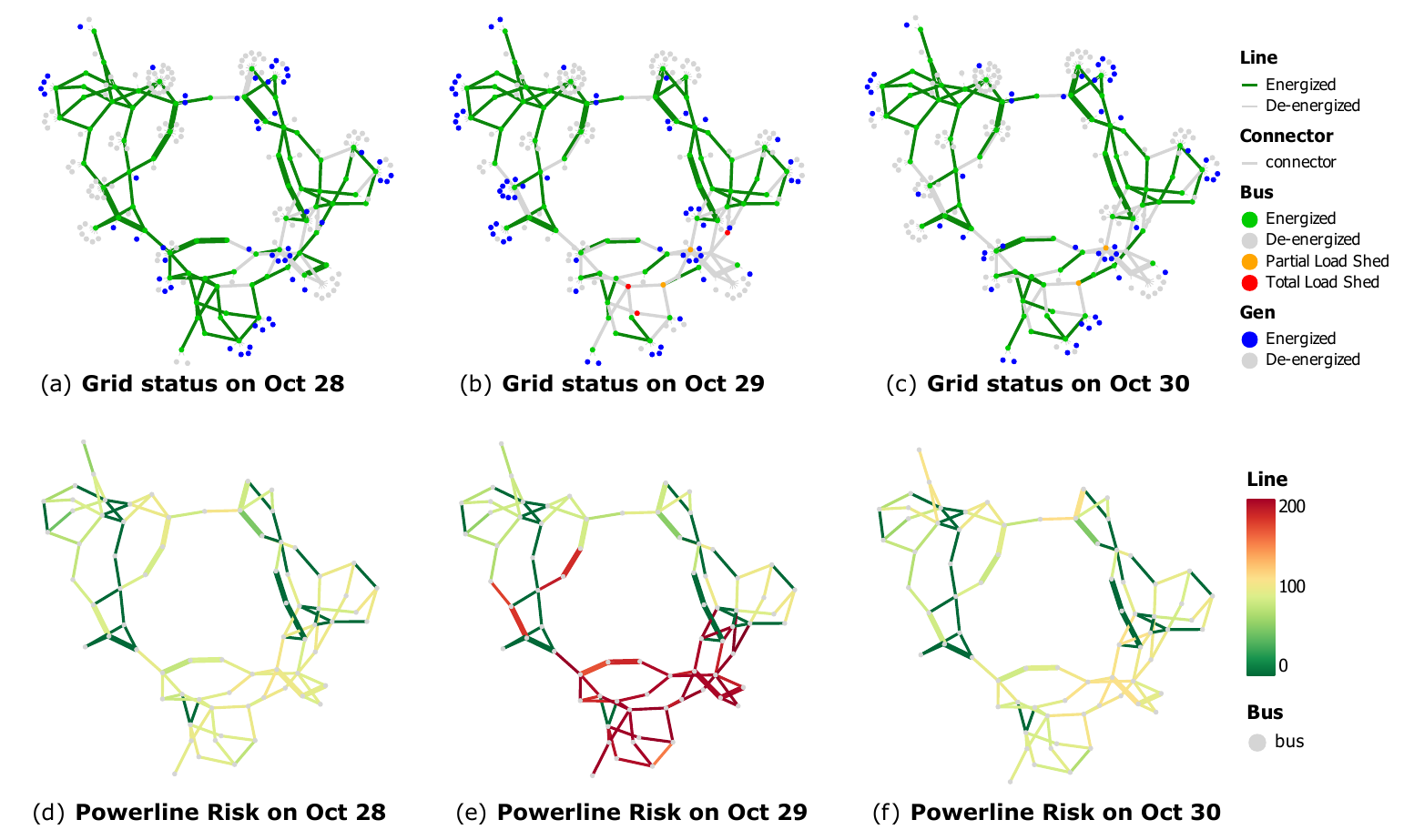}
    \caption{\small \textbf{Power grid status and risk levels from Oct 28-30.} Figs. \ref{fig:grid_risk}a-\ref{fig:grid_risk}c show the transmission lines status and Figs. \ref{fig:grid_risk}d-\ref{fig:grid_risk}f show the transmission line risk values for Oct 28 (left), 29 (middle) and 30 (right).}%
     \label{fig:grid_risk}
     \vspace{-5pt}
\end{figure*} 

\subsection{Baseline Solution}
We first test our method on the RTS-GMLC system using the baseline parameter settings. %

\subsubsection{Total system risk and load shed}
We first analyze the impact on total system risk and total load served for the baseline case.
Fig. \ref{fig:system_plot} (top) shows the total system risk value without (orange) and with (blue) power shutoffs, while Fig. \ref{fig:system_plot} (bottom) shows the total load without (orange) and with (blue) power shutoffs. 
In the first few days, the grid risk is moderate. A small number of lines are de-energized to reduce the risk, but no load shed occurs. On Oct 29th, a large spike in risk occurs, leading to a significant  grid shutoff, and resulting in around 5\% load shed for this day.  On Oct 30th, the wildfire risk values return to moderate levels, and many lines are restored to avoid significant load shed. 
However, the total system risk is still reduced because many lines remain de-energized.
\subsubsection{Geographical allocation of risk and load shed}
Next, we analyze the geographical locations of load shed and wildfire risk in the system during the high risk days.
Figure \ref{fig:grid_risk} shows the state of the grid (top) and the wildfire risk values for each line (bottom) on Oct 28, 29 and 30 (left to right).

We observe that on Oct 28 (left), there are several lines with risk values $R_{ijt}>\boldsymbol{V}$ that are de-energized, but
no load shed. %
On Oct 29 (middle), there is very high wildfire risk %
in the southern region of the grid, causing many of the lines in this region to be turned off and resulting in partial load shed at 2 buses and total load shed at 3 buses. %
Interestingly, we can see that some high risk lines remain energized to avoid further load shed. On Oct 30, the risk is reduced and many lines have been restored, but there is still partial load shed at 2 buses. %

\subsubsection{Solution time}
The baseline experiment involved solving the rolling horizon MOPSAR problem with a 4 day forecast horizon 22 times (for each of the 22 days), and solves in 29 seconds. However, the solve time is highly dependent on the problem parameters, in particular to the risk threshold $V$ and $\alpha$. In some instances (such as $V=0$, $\alpha=0.2$) it can take more than 14 hours to solve the problem. %

\subsection{Impact of Forecast Horizon}
We next investigate how the forecast horizon impacts solution time and solution quality by solving
the rolling horizon problem with forecast horizons ranging from 1 to 7 days. %
\subsubsection{Solution time}
The solution time increases from 1.93s for the 1-day horizon to 213.63s for the 7-day horizon (more than a 100 times increase). This indicates that the solve time may become prohibitively large for larger systems and long horizons, and highlights why it is important to understand how long the horizon needs to be to provide good solutions.

\begin{figure}[t]
    \centering
    \includegraphics[width=0.9\columnwidth]{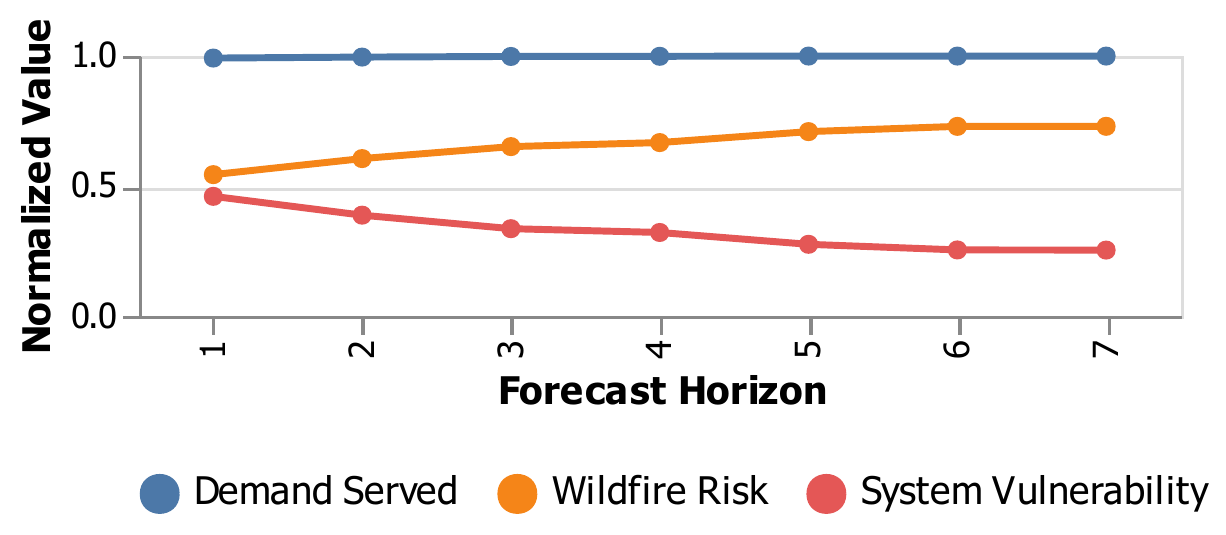}
    \caption{\small Objective value components for different forecast horizons.}
    \label{fig:s_horizon}
    \vspace{-15pt}
\end{figure}

\subsubsection{Objective function} %
We calculate the total load served, total system risk and total system vulnerability using the realized decisions and true wildfire risk for each time step and summing the values across the entire decision horizon. From these calculations, we found that the total objective function value improves by approximately 4.2\% as the forecast horizon increases from 1 to 7 days, with 3.2\% achieved already with a 3 day horizon. 
Figure \ref{fig:s_horizon} shows the different components of the objective function. %
The forecast horizon has a very small impact on the load shed, but the system risk increases and grid vulnerability decreases with a longer horizon.

\subsubsection{Line status}
We next assess how the forecast horizon changes line energization decisions.Table \ref{tab:horizon_sensitivity}(a) shows the total number of line energization and de-energization events that occur during the 21 day period. %
The number of de-energizations decreases from 93 de-energizations with a 1-day horizon to 74 de-energizations when using a 7-day horizon. The number and length of the restored lines show a less clear trend, with the number of re-energized lines remaining in the range from 61 to 54 across all horizons and the medium length horizon problems using a larger share of the restoration budget. %
Next, we consider the average, maximum and minimum risk level of energized and de-energized lines, shown in
Table \ref{tab:horizon_sensitivity}(b). %
All solutions keep some high risk lines (risk greater than the threshold $V=100$) in operation to serve load. Further, some low-risk lines (risk below the threshold $V=100$) are de-energized, as the restoration budget does not allow for restoration of all these lines. However, both the average and minimum line risk for de-energized lines is smaller for shorter forecast horizons, indicating that more low-risk lines (with risk below the threshold) remain de-energized due to an inadequate restoration budget and inability to plan further into the future. 

Overall, we conclude that a longer time horizon results in fewer de-energized lines because it better accounts for the slow restoration process, which may cause lines to remain de-energized after the risk has decreased, thus incurring future load shed and high vulnerability penalties.

\begin{table}[t]
    \centering
    \caption{Results for varying forecast horizons.}
    \label{tab:horizon_sensitivity}
    \begin{subtable}{\columnwidth}
        \centering
        \caption{De-energization and Re-energization Decisions}
        \vspace{-2pt}
        \resizebox{\columnwidth}{!}
        {\begin{tabular}{l|ccccccc}
\hline 
\hline 
\textbf{Horizon (days)}& \textbf{1} & \textbf{2} & \textbf{3} & \textbf{4} & \textbf{5} & \textbf{6} & \textbf{7} \\ \hline
\hline
\textbf{De-energization} & & & & & & & \\
\textbf{   \# Lines}& 93 & 85 & 84 & 85 & 79 & 78 & 74 \\ 
\hline
\hline
\textbf{Re-energization} & & & & & & & \\
\textbf{   \# Lines}& 61 & 58 & 61 & 60 & 57 & 58 & 54 \\
\textbf{   \# Miles}& 1316 & 1346 & 1436 & 1381 & 1385 & 1373 & 1307 \\
\hline
\hline 

\end{tabular}
}
        \label{tab:h_activations}
    \end{subtable}
    \begin{subtable}{\columnwidth}
        \centering
        \vspace{1em}
        \caption{Wildfire risk of Energized and De-energized Lines}
        \vspace{-2pt}
        \resizebox{\columnwidth}{!}
        {\begin{tabular}{lr|ccccccc}
\hline
\hline
\multicolumn{2}{l|}{\textbf{Horizon (days)}} \rule{0pt}{2ex}& \textbf{1} & \textbf{2} & \textbf{3} & \textbf{4} & \textbf{5} & \textbf{6} & \textbf{7} \\ \hline
\hline
\multicolumn{1}{l}{\multirow{3}{*}{\begin{tabular}[c]{@{}l@{}}\textbf{Energized}\\ \textbf{Line Risk}\end{tabular}}} \rule{0pt}{2ex} & \textbf{Avg}& 50 & 52 & 54 & 55 & 56 & 57 & 57 \\
\multicolumn{1}{l}{} & \textbf{Min}& 0 & 0 & 0 & 0 & 0 & 0 & 0 \\
\multicolumn{1}{l}{} & \textbf{Max}& 198 & 208 & 206 & 206 & 208 & 208 & 208 \\ 
\hline
\hline
\multicolumn{1}{l}{\multirow{3}{*}{\begin{tabular}[c]{@{}l@{}}\textbf{De-energized}\\ \textbf{Line Risk}\end{tabular}}} \rule{0pt}{2ex} & \textbf{Avg}& 100 & 102 & 104 & 104 & 106 & 107 & 107 \\
\multicolumn{1}{l}{} & \textbf{Min}& 1 & 1 & 17 & 8 & 40 & 40 & 40 \\
\multicolumn{1}{l}{} & \textbf{Max}& 215 & 215 & 215 & 215 & 215 & 215 & 215 \\ 
\hline
\hline
\end{tabular}
}
        \label{tab:h_risk}
        \vspace{-5pt}
    \end{subtable}
        \vspace{-5pt}
\end{table}

\subsection{Impact of Restoration Budget}
We next investigate the impact of the restoration budget $\boldsymbol{Y}_t$. %
\subsubsection{Objective value} Table \ref{tab:restoration_sensitivity}(a) shows that the objective function value improves as the restoration budget increases. This is as expected, as a higher budget relaxes the problem. We further found that the load shed remains similar across all restoration budgets, while the wildfire risk increases and the vulnerability decreases with higher budgets. 

\subsubsection{Line status}
Table \ref{tab:restoration_sensitivity}(b) shows %
the number of lines that were de-energized and re-energized, along with the total and average length of the restored lines. %
We observe that a larger restoration budget significantly increases both the number of de-energized and re-energized lines. Further, solutions with a small restoration budget de-energize many more lines than they re-energize, leaving a large number of lines still de-energized at the end of the study period. 
The average length of the restored lines %
increases from 11 miles/line with the lowest restoration to 28 miles/line with the highest restoration budget. This shows that the system operator is able to re-energize longer lines if more restoration capacity is available.

Table \ref{tab:restoration_sensitivity}(c) shows that the average, minimum and maximum wildfire risk of transmission lines change as we increase the restoration budget. %
The maximum risk of energized lines decreases and the minimum risk of de-energized lines increases as the restoration budget becomes higher. This is because the operator is able to react faster to changes in the wildfire risk when we have a higher restoration budget.

\begin{table}[t]
    \centering
    \caption{Results for different restoration budgets.}
    \begin{subtable}{\columnwidth}
        \centering
        \caption{Objective Function Values} %
        \vspace{-2pt}
        \resizebox{\columnwidth}{!}
        {
\begin{tabular}{r|ccccc}
\hline
\hline
\rule{0pt}{2ex}
\textbf{Restoration Budget}& \textbf{25} & \textbf{50} & \textbf{75} & \textbf{100} & \textbf{125} \\
\hline
\hline
\textbf{Objective value}& -0.402 & -0.395 & -0.391 & -0.388 & -0.385 \rule{0pt}{2ex}\\
\textbf{\% Change}& 0\% & +1.74\% & +1.77\% & +3.48\% & +4.23\% \rule{0pt}{2ex}
\\ \hline \hline
\end{tabular}
}
        \label{tab:r_obj}
    \end{subtable}
    \begin{subtable}{\columnwidth}
        \centering
        \vspace{1em}
        \caption{De-energization and Re-energization Decisions}
        \vspace{-2pt}
        {
\begin{tabular}{l|ccccc}
\hline 
\hline \textbf{Restoration Budget}& \textbf{25} & \textbf{50} & \textbf{75} & \textbf{100} & \textbf{125} \\ 
\textbf{Total Budget}& 550 & 1100 & 1650 & 2200 & 2750 \\
\hline
\hline
\textbf{De-energization} & & & & & \\
\textbf{   \# Lines}& 66 & 77 & 85 & 95 & 99  \\ 
\hline
\hline
\textbf{Re-energization} & & & & & \\
\textbf{   \# Lines}& 28 & 45 & 60 & 75 & 83  \\
\textbf{   \# Miles}& 308 & 890 & 1381 & 1907 & 2325 \\
\textbf{Avg. length (Miles)} & 11 & 19.8 & 23.0 & 25.4 & 28.0\\
\hline
\hline 

\end{tabular}
}
        \label{tab:r_activations}
    \end{subtable}
    \begin{subtable}{\columnwidth}
        \centering
        \vspace{1em}
        \caption{Wildfire risk of Energized and De-energized Lines}
        \vspace{-2pt}
        {\begin{tabular}{lr|ccccc}
\hline
\hline
\multicolumn{2}{l|}{\textbf{Restoration Budget}} \rule{0pt}{2ex}& \textbf{25} & \textbf{50} & \textbf{75} & \textbf{100} & \textbf{125} \\ \hline \hline
\multicolumn{1}{l}{\multirow{3}{*}{\begin{tabular}[c]{@{}l@{}}\textbf{Energized} \\ \textbf{Line Risk}\end{tabular}}} \rule{0pt}{2ex} & \textbf{Avg}& 54 & 54 & 55 & 55 & 56 \\
\multicolumn{1}{l}{} & \textbf{Min}& 0 & 0 & 0 & 0 & 0 \\
\multicolumn{1}{l}{} & \textbf{Max}& 208 & 208 & 206 & 206 & 206 \\ \hline \hline
\multicolumn{1}{l}{\multirow{3}{*}{\begin{tabular}[c]{@{}l@{}}\textbf{De-energized} \\ \textbf{Line Risk}\end{tabular}}} \rule{0pt}{2ex} & \textbf{Avg}& 100 & 103 & 104 & 107 & 108 \\
\multicolumn{1}{l}{} & \textbf{Min}& 1 & 8 & 8 & 40 & 48 \\
\multicolumn{1}{l}{} & \textbf{Max}& 215 & 215 & 215 & 215 & 215 \\ \hline \hline
\end{tabular}
}
        \label{tab:r_risk}
    \end{subtable}
    \label{tab:restoration_sensitivity}
        \vspace{-10pt}
\end{table}

\subsection{Impact of Risk Threshold}
\begin{figure}[t]
    \centering
    \begin{subfigure}{0.33\textwidth}
        \centering
        \includegraphics[width=0.9\textwidth]{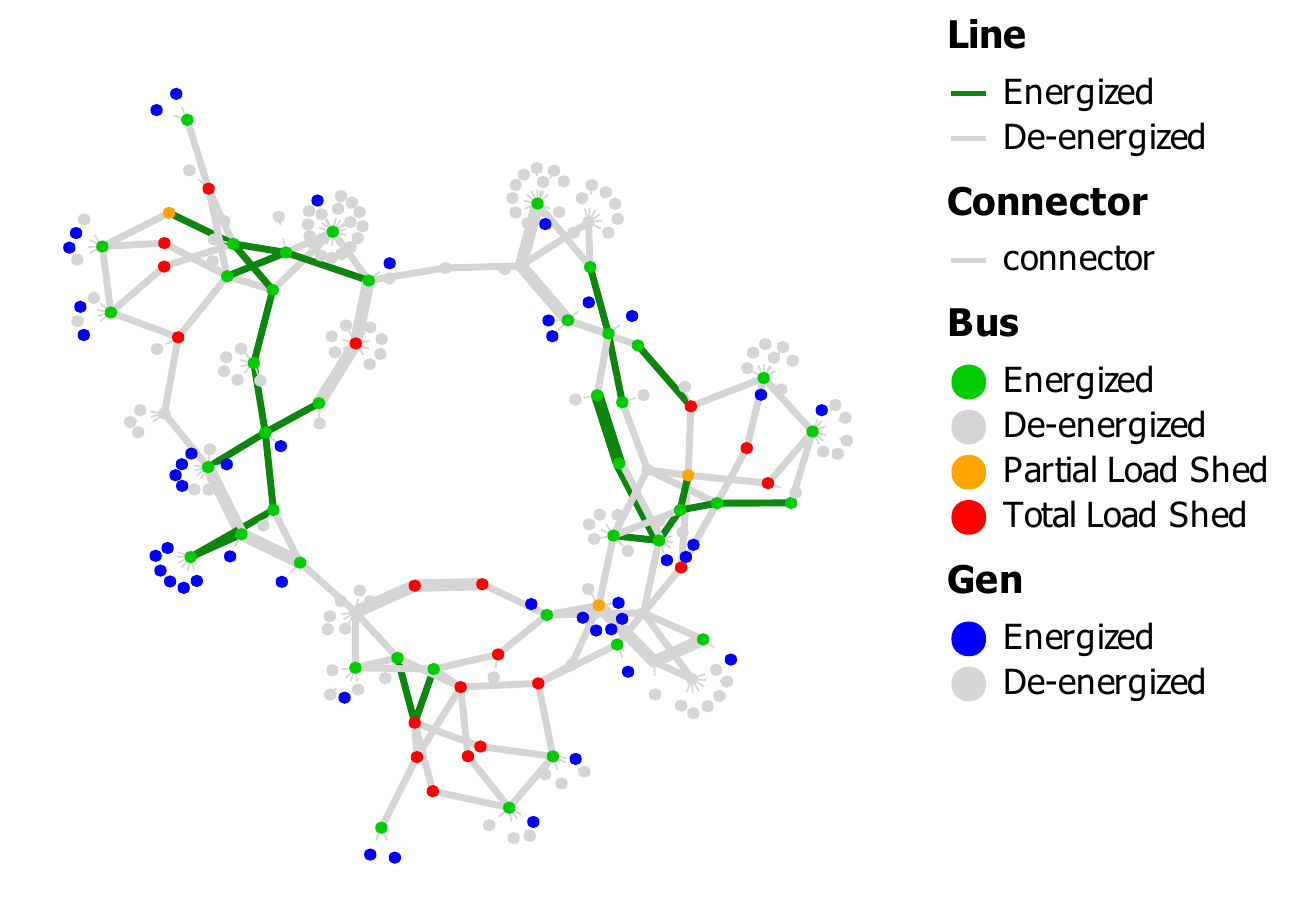} 
        \caption{\footnotesize $\boldsymbol{V}=0, \; \boldsymbol{\alpha}=0.7, \; D_{served}=1413$} \label{fig:vul0}
     \end{subfigure}
         \begin{subfigure}{0.33\textwidth}
        \centering
        \includegraphics[width=0.9\textwidth]{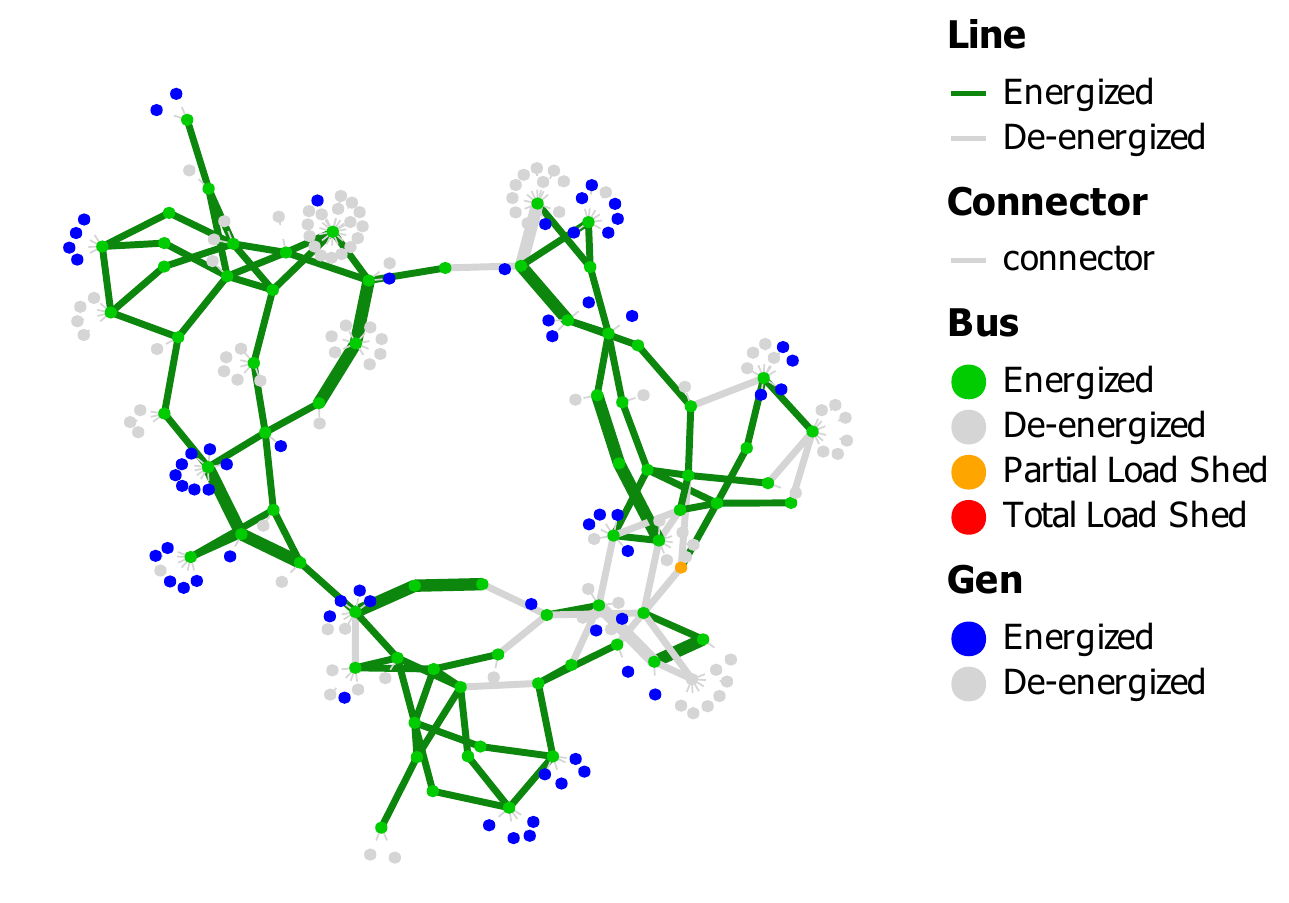} 
        \caption{\footnotesize $\boldsymbol{V}=100, \; \boldsymbol{\alpha}=0.7, \; D_{served}=2038$} \label{fig:vul100}
    \end{subfigure}
    \caption{\small System status on Oct 29th for two values of $\boldsymbol{V}=0$ (top) and $\boldsymbol{V}=100$ (bottom).} %
     \label{fig:vul_plot}
     \vspace{-15pt}
\end{figure}

To demonstrate the benefit of including a risk threshold that penalizes the shutoff of low-risk lines, we compare the solutions for %
two different risk thresholds $\boldsymbol{V}=0$ (no penalty for shutoff of low-risk lines) and $\boldsymbol{V}=100$ (baseline).
Figure \ref{fig:vul_plot}(a) and (b) show the network status on October 29th, the highest risk day, for each of the two thresholds, respectively. %
The solution with risk threshold $\boldsymbol{V}=0$ turns off even low risk lines, leading to many de-energized lines and significant load shed. In comparison, the solution with $\boldsymbol{V}=100$ only disables a few lines and %
maintains some redundancy in the network. The two solutions serve 1413 MW and 2038 MW of load and have a wildfire risk of 4,920 and 114,220, respectively. We thus conclude that a non-zero risk threshold promotes more system redundancy and higher load served for a given value of $\boldsymbol{\alpha}$, but also has significantly higher wildfire risk.

\section{CONCLUSION} \label{sec:conclusion}

This paper proposes the Multi-period Optimal Power Shutoff And Restoration (MOPSAR) problem, which co-optimizes preemptive power shutoffs and restoration efforts in power systems with high wildfire risk. %
The MOPSAR problem is implemented in a rolling horizon framework where the schedules are re-optimized on a daily basis as new information about wildfire risk become available. %
We apply the proposed method to the RTS-GMLC system and  find that 
\emph{(i)} a longer forecast horizon better accounts for the impact of de-energization on load shed, 
\emph{(ii)} a larger restoration budget reduces risk by allowing for disabling and restoring more lines, 
and \emph{(iii)} a higher risk threshold incentivizes more system redundancy, thus increasing operational security but also wildfire risk. %

The proposed framework provide several avenues for future work. 
For example, there is a need to extend the model to consider N-1 security constraints
and investigate the impact of using the full AC power flow model. 
Further, more efficient
solution techniques are necessary to plan power shutoffs on a realistic scale power network.

\section{ACKNOWLEDGMENTS}
The authors would like to acknowledge Sofia Taylor for helping obtain wildfire risk data for the RTS-GMLC system. %

\Urlmuskip=0mu plus 1mu\relax
\bibliographystyle{IEEEtran}
\bibliography{IEEEabrv,references}

\begin{thebibliography}{10}
\providecommand{\url}[1]{#1}
\csname url@samestyle\endcsname
\providecommand{\newblock}{\relax}
\providecommand{\bibinfo}[2]{#2}
\providecommand{\BIBentrySTDinterwordspacing}{\spaceskip=0pt\relax}
\providecommand{\BIBentryALTinterwordstretchfactor}{4}
\providecommand{\BIBentryALTinterwordspacing}{\spaceskip=\fontdimen2\font plus
\BIBentryALTinterwordstretchfactor\fontdimen3\font minus
  \fontdimen4\font\relax}
\providecommand{\BIBforeignlanguage}[2]{{%
\expandafter\ifx\csname l@#1\endcsname\relax
\typeout{** WARNING: IEEEtran.bst: No hyphenation pattern has been}%
\typeout{** loaded for the language `#1'. Using the pattern for}%
\typeout{** the default language instead.}%
\else
\language=\csname l@#1\endcsname
\fi
#2}}
\providecommand{\BIBdecl}{\relax}
\BIBdecl

\bibitem{2009_victorian}
\emph{\BIBforeignlanguage{en}{Final report}}.\hskip 1em plus 0.5em minus
  0.4em\relax Melbourne: Parliament of Victoria, 2009 Victorian Bushfires Royal
  Commission, 2010.

\bibitem{noauthor_stats_nodate}
\BIBentryALTinterwordspacing
``Stats \& {Events}.'' [Online]. Available:
  \url{https://www.fire.ca.gov/stats-events/}
\BIBentrySTDinterwordspacing

\bibitem{7995099}
D.~N. Trakas and N.~D. Hatziargyriou, ``Optimal distribution system operation
  for enhancing resilience against wildfires,'' \emph{IEEE Trans. on Power
  Systems}, vol.~33, no.~2, pp. 2260--2271, 2018.

\bibitem{CHOOBINEH201520}
M.~Choobineh, B.~Ansari, and S.~Mohagheghi, ``Vulnerability assessment of the
  power grid against progressing wildfires,'' \emph{Fire Safety Journal},
  vol.~73, pp. 20--28, 2015.

\bibitem{mohagheghi2015optimal}
S.~Mohagheghi and S.~Rebennack, ``Optimal resilient power grid operation during
  the course of a progressing wildfire,'' \emph{International Journal of
  Electrical Power \& Energy Systems}, vol.~73, pp. 843--852, 2015.

\bibitem{moutis2022pmu}
P.~Moutis and U.~Sriram, ``Pmu-driven non-preemptive disconnection of overhead
  lines at the approach or break-out of forest fires,'' \emph{IEEE Trans. on
  Power Systems}, 2022.

\bibitem{nazemi2022powering}
M.~Nazemi and P.~Dehghanian, ``Powering through wildfires: An integrated
  solution for enhanced safety and resilience in power grids,'' \emph{IEEE
  Trans. on Industry Applications}, vol.~58, no.~3, pp. 4192--4202, 2022.

\bibitem{8768218}
S.~Jazebi, F.~de~León, and A.~Nelson, ``Review of wildfire management
  techniques—part i: Causes, prevention, detection, suppression, and data
  analytics,'' \emph{IEEE Trans. on Power Delivery}, vol.~35, no.~1, pp.
  430--439, 2020.

\bibitem{8767948}
------, ``Review of wildfire management techniques—part ii: Urgent call for
  investment in research and development of preventative solutions,''
  \emph{IEEE Trans. on Power Delivery}, vol.~35, no.~1, pp. 440--450, 2020.

\bibitem{arab2021three}
A.~Arab, A.~Khodaei, R.~Eskandarpour, M.~P. Thompson, and Y.~Wei, ``Three lines
  of defense for wildfire risk management in electric power grids: A review,''
  \emph{IEEE Access}, 2021.

\bibitem{muhs2020wildfire}
J.~W. Muhs, M.~Parvania, and M.~Shahidehpour, ``Wildfire risk mitigation: A
  paradigm shift in power systems planning and operation,'' \emph{IEEE Open
  Access Journal of Power and Energy}, vol.~7, pp. 366--375, 2020.

\bibitem{noauthor_pge_nodate}
\BIBentryALTinterwordspacing
``{PG}\&{E} {Announces} {Major} {New} {Electric} {Infrastructure} {Safety}
  {Initiative} to {Protect} {Communities} {From} {Wildfire} {Threat}.''
  [Online]. Available: \url{https://tinyurl.com/5bzkty6d}
\BIBentrySTDinterwordspacing

\bibitem{noauthor_psps_nodate}
\BIBentryALTinterwordspacing
``{Power} {Restored} for {Essentially} {All} {Affected} {Customers} {After}
  {Dry}, {Offshore} {Wind} {Event} and {Exceptional} {Drought} {Conditions}
  {Prompt} {Safety} {Shutoff}.'' [Online]. Available:
  \url{https://tinyurl.com/2v8ajfum}
\BIBentrySTDinterwordspacing

\bibitem{noauthor_utility_nodate}
\BIBentryALTinterwordspacing
``Utility {Company} {PSPS} {Post} {Event} {Reports}.'' [Online]. Available:
  \url{https://www.cpuc.ca.gov/consumer-support/psps/utility-company-psps-post-event-reports}
\BIBentrySTDinterwordspacing

\bibitem{rhodes2020balancing}
N.~Rhodes, L.~Ntaimo, and L.~Roald, ``Balancing wildfire risk and power outages
  through optimized power shut-offs,'' \emph{IEEE Trans. on Power Systems},
  vol.~36, no.~4, pp. 3118--3128, 2020.

\bibitem{hong2022data}
W.~Hong, B.~Wang, M.~Yao, D.~Callaway, L.~Dale, and C.~Huang, ``Data-driven
  power system optimal decision making strategy underwildfire events,''
  Lawrence Livermore National Lab.(LLNL), Livermore, CA (United States), Tech.
  Rep., 2022.

\bibitem{umunnakwe2022data}
A.~Umunnakwe, M.~Parvania, H.~Nguyen, J.~D. Horel, and K.~R. Davis,
  ``Data-driven spatio-temporal analysis of wildfire risk to power systems
  operation,'' \emph{IET Generation, Transmission \& Distribution}, vol.~16,
  no.~13, pp. 2531--2546, 2022.

\bibitem{bayani2022quantifying}
R.~Bayani, M.~Waseem, S.~D. Manshadi, and H.~Davani, ``Quantifying the risk of
  wildfire ignition by power lines under extreme weather conditions,''
  \emph{IEEE Systems Journal}, 2022.

\bibitem{lesage2022optimally}
A.~Lesage-Landry, F.~Pellerin, J.~A. Taylor, and D.~S. Callaway, ``Optimally
  scheduling public safety power shutoffs,'' \emph{arXiv preprint
  arXiv:2203.02861}, 2022.

\bibitem{kody2022optimizing}
A.~Kody, R.~Piansky, and D.~K. Molzahn, ``Optimizing transmission
  infrastructure investments to support line de-energization for mitigating
  wildfire ignition risk,'' \emph{arXiv preprint arXiv:2203.10176}, 2022.

\bibitem{yang2022resilient}
W.~Yang, S.~N. Sparrow, M.~Ashtine, D.~C. Wallom, and T.~Morstyn, ``Resilient
  by design: Preventing wildfires and blackouts with microgrids,''
  \emph{Applied Energy}, vol. 313, p. 118793, 2022.

\bibitem{astudillo2022managing}
A.~Astudillo, B.~Cui, and A.~S. Zamzam, ``Managing power systems-induced
  wildfire risks using optimal scheduled shutoffs,'' National Renewable Energy
  Lab, Golden, CO (United States), Tech. Rep., 2022.

\bibitem{gorka2022efficient}
J.~Gorka and L.~Roald, ``Efficient representations of radiality constraints in
  optimization of islanding and de-energization in distribution grids,''
  \emph{Power Systems Computational Conference 2022}, 2022.

\bibitem{kody2022sharing}
A.~Kody, A.~West, and D.~K. Molzahn, ``Sharing the load: Considering fairness
  in de-energization scheduling to mitigate wildfire ignition risk using
  rolling optimization,'' \emph{arXiv preprint arXiv:2204.06543}, 2022.

\bibitem{yao2022predicting}
M.~Yao, M.~Bharadwaj, Z.~Zheng, B.~Jin, and D.~S. Callaway, ``Predicting
  electricity infrastructure induced wildfire risk in california,''
  \emph{Environmental Research Letters}, 2022.

\bibitem{pgeOct2020report}
\BIBentryALTinterwordspacing
``Pacific gas and electric company public safety power shutoff (psps) report to
  the cpuc october 25-28, 2020 de-energization event,'' Pacific Gas \&
  Electric. [Online]. Available: \url{https://tinyurl.com/38e8ykky}
\BIBentrySTDinterwordspacing

\bibitem{rhodes2021powermodelsrestoration}
N.~Rhodes, D.~M. Fobes, C.~Coffrin, and L.~Roald, ``Powermodelsrestoration. jl:
  An open-source framework for exploring power network restoration
  algorithms,'' \emph{Electric Power Systems Research}, vol. 190, p. 106736,
  2021.

\bibitem{wfpi}
\BIBentryALTinterwordspacing
``Wildland fire potential index (wfpi).'' [Online]. Available:
  \url{https://www.usgs.gov/fire-danger-forecast/wildland-fire-potential-index-wfpi}
\BIBentrySTDinterwordspacing

\bibitem{Burak2016}
\BIBentryALTinterwordspacing
B.~Kocuk, H.~Jeon, S.~S. Dey, J.~Linderoth, J.~Luedtke, and X.~A. Sun, ``A
  cycle-based formulation and valid inequalities for dc power transmission
  problems with switching,'' \emph{Oper. Res.}, vol.~64, no.~4, p. 922–938,
  Aug. 2016. [Online]. Available:
  \url{https://doi-org.ezproxy.library.wisc.edu/10.1287/opre.2015.1471}
\BIBentrySTDinterwordspacing

\bibitem{julia}
\BIBentryALTinterwordspacing
J.~Bezanson, A.~Edelman, S.~Karpinski, and V.~Shah, ``Julia: A fresh approach
  to numerical computing,'' \emph{SIAM Review}, vol.~59, no.~1, pp. 65--98,
  2017. [Online]. Available: \url{https://doi.org/10.1137/141000671}
\BIBentrySTDinterwordspacing

\bibitem{gurobi}
{Gurobi Optimization, Inc.}, ``Gurobi optimizer reference manual,'' Published
  online at \url{http://www.gurobi.com}, 2014.

\bibitem{RTSGMLC}
C.~Barrows, A.~Bloom, A.~Ehlen, J.~Ikäheimo, J.~Jorgenson, D.~Krishnamurthy,
  J.~Lau, B.~McBennett, M.~O’Connell, E.~Preston, A.~Staid, G.~Stephen, and
  J.-P. Watson, ``The ieee reliability test system: A proposed 2019 update,''
  \emph{IEEE Trans. on Power Systems}, vol.~35, no.~1, pp. 119--127, 2020.

\bibitem{pglib}
{PGLib Optimal Power Flow Benchmarks}, ``The ieee pes task force on benchmarks
  for validation of emerging power system algorithms,'' Published online at
  \url{https://github.com/power-grid-lib/pglib-opf}, 2021.

\bibitem{taylor2022framework}
S.~Taylor and L.~A. Roald, ``A framework for risk assessment and optimal line
  upgrade selection to mitigate wildfire risk,'' \emph{Electric Power Systems
  Research}, vol. 213, p. 108592, 2022.

\end{thebibliography}

\end{document}